\documentclass[a4paper,11pt]{article}
\pdfoutput=1
\usepackage{jheppub} 
\usepackage{graphicx,color,rotating}
\usepackage{hyperref}
\usepackage{epsfig,color}
\usepackage{slashed}
\usepackage{amsfonts}
\usepackage{ulem}

\newcommand{\gsim}{\raisebox{-0.13cm}{~\shortstack{$>$ \\[-0.07cm] $\sim$}}~}

\begin{document}
\title{Stellar cooling, inelastic dark matter, and XENON}

\renewcommand{\thefootnote}{\arabic{footnote}}

\author{
Wai-Yee Keung$^{1}$,  Danny Marfatia$^{2}$, and
Po-Yan Tseng$^{3}$}
\affiliation{
$^1$ Department of Physics, University of Illinois at Chicago,
Illinois 60607 USA \\
$^2$ Department of Physics \& Astronomy, University of Hawaii at Manoa,
Honolulu, HI 96822, USA \\
$^3$ Department of Physics and IPAP, Yonsei University,
Seoul 03722, Republic of Korea \\
}
\date{\today}

\abstract{
We consider a novel scenario of dark photon-mediated inelastic dark matter to explain the white dwarf cooling 
excess suggested by its luminosity function, and the
excess in electron recoil events at XENON1T. In the Sun, the dark photon $A'$ is produced mainly via
thermal processes, and 
the heavier dark matter $\chi_2$ 
is produced by the scattering of halo dark matter $\chi_1$ with electrons. 
The XENON1T signal arises primarily by solar $A'$ scattering, and $A'$ emission by white dwarfs accommodates the extra cooling while maintaining
consistency with other stellar cooling observations. A tritium component in the XENON1T detector is also required.
 We show for parameters that explain the XENON1T data, but not the white dwarf cooling anomaly, that
 a second signal peak may be buried in the XENON1T data and revealable at XENONnT. However, the parameters
 that give the double peak in the spectrum are incompatible with constraints from horizontal branch stars.

}

\maketitle

\section{Introduction}

Astrophysical  observations of
several stellar systems including white dwarfs (WDs)~\cite{kepler,Isern:1992gia,Corsico:2012ki,Corsico:2012sh,Bertolami:2014wua,Bertolami:2014noa} and horizontal branch (HB) stars~\cite{Ayala:2014pea}
show evidence of excess cooling in comparison 
to standard theoretical predictions. The number density of WDs as a function of brightness, called the white dwarf luminosity function (WDLF),
indicates a cooling anomaly at the $4\sigma$ level~\cite{Giannotti:2015kwo}. Pulsating WDs with a DA spectral type also show a
cooling anomaly~\cite{Corsico:2012ki,Corsico:2012sh}.
%
%
The measured value of the $R$-parameter, which is the ratio of the number of stars in the HB 
to that in the upper portion of the RGB (Red Giant Branch) in globular clusters,
is smaller than predicted, hinting that HB stars cool more efficiently than expected~\cite{Ayala:2014pea}.
%
However, HB stars (aside from the aforementioned hints), RGB stars (because of their higher core temperatures) and the Sun (whose
observed luminosity is consistent with the standard solar model) restrict
excess cooling. 

Recently, the XENON1T experiment with its 1024~kg fiducial volume
and 0.65~ton-year exposure of xenon~\cite{Aprile:2020tmw}
reported a $3\sigma$ excess in the electron recoil spectrum
between 2-3~keV.
Various explanations have been put forward~\cite{coredump}. 

We propose a joint explanation of the WDLF cooling and XENON1T excesses in the context of
inelastic dark matter (DM)~\cite{TuckerSmith:2001hy} mediated by dark photons~\cite{Batell:2009vb}; for similar scenarios  
in connection with XENON1T data see Refs.~\cite{Alonso-Alvarez:2020cdv,Harigaya:2020ckz,Baryakhtar:2020rwy,Bramante:2020zos,Bloch:2020uzh}.
We consider two relic Majorana fermion DM particles $\chi_1$ and $\chi_2$ of MeV mass
that form a pseudo-Dirac pair with a small mass gap,
$\Delta m_\chi \equiv m_{\chi_2} - m_{\chi_1}  \approx $~keV. A 
dark photon $A'$ of sub-GeV mass mediates the interaction  
 between $\chi_1$ and $\chi_2$. 
The DM halo is constituted primarily by $\chi_1$ for the
parameter values of our scenario~\cite{Bramante:2020zos}.
These new particles connect the  
WDLF anomaly and XENON1T excess.
%
Both $\chi_2$ and $A'$ can be produced in
the thermal plasma of the Sun, and 
contribute to the XENON1T signal. 
The $\chi_2$ in the Sun are created by the relic $\chi_1$ through 
the process $\chi_1 e \to \chi_2 e$.
$A'$ production is dominated by bremsstrahlung~\cite{An:2013yfc,Redondo:2013lna}.
After propagation from the Sun to Earth, the down scattering
$\chi_2 e \to \chi_1 e$ and absorption $A'e \to e$ processes
yield the electron recoil signals in XENON1T.
We find a region of parameter space favored by both the WDLF anomaly
and XENON1T excess, and that is consistent with constraints from
other stellar cooling observations, provided a small contribution from the $\beta$-decay of tritium is present. It is noteworthy that scenarios invoking solar axion-like particles
to explain the XENON1T excess are not consistent with constraints from stellar cooling, 
especially the $R$-parameter~\cite{DiLuzio:2020jjp}.

The paper is organized as follows.
In section~\ref{sec:model}, 
we study our dark photon-mediated inelastic DM scenario and demonstrate that if
$m_{A'} < \Delta m_\chi, 10$~keV both excesses can be explained.
In section~\ref{sec:XENONnT}, we show that a
distinctive double-peak signal may be present in XENON1T data  
and that can be confirmed by XENONnT.
We summarize in section~\ref{sec:conclusion}.

\bigskip

\section{Dark photon-mediated inelastic dark matter}
\label{sec:model}

$A'$ couples to $\chi_1,\chi_2$ and the electron
via the effective interactions,
\begin{eqnarray}
\label{eq:scenario1_interaction}
\mathcal{L}\supset (\epsilon e) A'_\mu(\bar{e}\gamma^\mu e)
+\left( \frac{i\,g_\chi}{2}  A'_\mu (\bar{\chi_2}\gamma^\mu \chi_1) + {\rm h.c.}  \right)\,,
\end{eqnarray}
where the $A'ee$ coupling may originate from the kinetic mixing 
between $A'$ and the photon via 
$-\frac{\epsilon}{2}F'_{\mu\nu}F^{\mu\nu}$~\cite{Holdom:1985ag}.
%
Both $A'_\mu(\bar{\chi_2}\gamma^\mu \chi_1)$ 
and its Hermitian conjugate contribute to the XENON1T signal
due to their Majorana property.
We insert the imaginary unit $i$ so that  $g_\chi$ is a real number.

\subsection{\boldmath{$\Delta m_\chi\,, 40~{\rm keV} < m_{A'}$}}

Since $m_{A'}> \Delta m_\chi$,  the 2-body decay $\chi_2 \to A' \chi_1$ 
is kinematically forbidden, and
$\chi_2$ is stable on the length scale of the solar system. 
The signal in XENON1T data is produced by the down scattering  process,
\begin{eqnarray}
\chi_2+e \to \chi_1 + e\,.
\end{eqnarray}
We assume
$\Delta m_\chi \simeq 3~{\rm keV}$
to obtain a peak in the XENON1T electron recoil spectrum at around 3~keV~\cite{Aprile:2020tmw}.
The central region of the Sun which has a keV temperature electron plasma excites $\chi_1$ from the DM halo to produce a flux of $\chi_2$ 
via 
\begin{eqnarray}
\chi_1+ e \to \chi_2 + e\,.
\end{eqnarray}
The rate for this process is given by
\cite{Dasgupta:2020dik,McDermott:2011jp}
\begin{eqnarray}
C_c &=& \frac{\rho_{\chi_1}}{m_{\chi_1}}\, \sum_{i=1,2,3} N^{i}_e
\int du \frac{f_{\rm DM}(u)}{u}\left(u^2+(v^i_{\rm esc})^2\right) g_1(u) 
 \int f_e(K_e,T^i) \sigma_{\chi_1 e \to \chi_2 e}\,  dK_e 
\,,
\end{eqnarray}
where we divide the Sun into three shells with $i=1,2,3$ corresponding to $0\leq R<0.1R_\odot$, $0.1R_\odot\leq R<0.18R_\odot$, 
and $0.18R_\odot\leq R<0.35R_\odot$, respectively,
and sum over the contributions. 
Here, $N^i_e\simeq 9\times 10^{55}\,,2.1\times 10^{56}\,,4.7\times 10^{56}$
are the number of electrons in each shell and $v^i_{\rm esc} (\rm {km/s})=1335,1226,1040$ are the corresponding escape velocities from the outer
surface of each shell,
$\rho_{\chi_1}=0.4~{\rm GeV/cm^3}$ is the local DM density,{\footnote{Absent a concrete model, we do not attempt to estimate the
DM relic abundance from thermal freeze-out, and allow for non-thermal production to be determinative.} and
$f_{\rm DM}$ is the Maxwell-Boltzmann velocity distribution of DM with dispersion 270~km/s. 
The capture probability $g_1(u)$ depends on the energy loss distribution $(1/\sigma) d\sigma/d\Omega_{\rm CM}$, which need not be
uniform if the mediator is lighter than either of the scattered particles~\cite{Dasgupta:2020dik}. 
The Boltzmann distribution of electrons with kinetic energy $K_e$ is
$$
f_e(K_e,T^i)=2\sqrt{\frac{K_e}{\pi}}\,
\left( \frac{1}{T^i} \right)^{3/2} 
e^{-K_e/T^i}\,,
$$
where $T^i/{\rm keV} \simeq 1.21,1.02,0.74$ are the average temperatures in each shell. The contribution to the capture rate for radii
larger than $0.35R_\odot$ is negligible because of the falling temperature and  exponentially falling electron number density.
The total rate cannot exceed the geometric limit for the solar region within $0.35R_\odot$~\cite{Garani:2018kkd},
\begin{eqnarray}
C^\odot|_{\rm geom}=\frac{5.4\times 10^{29}}{\rm s}
~\frac{\rho_{\chi_1}}{\rm GeV/cm^3}
~\frac{\rm GeV}{m_{\chi_1}}
\,,
\end{eqnarray}
because in this limit all DM particles within the geometric area of the solar core are captured.
For $C_c > C^\odot|_{\rm geom}$, we take $C_c = C^\odot|_{\rm geom}$.

The amplitude squared for $e(p_1)+\chi_1(p_2) \to e(p_3)+\chi_2(p_4)$ is given by
\begin{eqnarray}
\frac{1}{4} \sum |M|^2 &=& \frac{(\epsilon e g_\chi)^2}{(t-m^2_{A'})^2} \left\lbrace 
2t^2+4st+4s^2-2(t+2s)(m^2_{\chi_1}+m^2_{\chi_2}) 
\right. \nonumber \\
&& \left.
+4tm_{\chi_1}m_{\chi_2}+4m^2_{\chi_1}m^2_{\chi_2}
+4m^2_e(m^2_e-2s+2m_{\chi_1}m_{\chi_2})
\right\rbrace\,.
\end{eqnarray}
In the $\chi_1$ rest frame, we define $K_e$ and $K_{\chi_2}$ to be the kinetic energy of the incoming electron and outgoing $\chi_2$, respectively.
Then, the Mandelstam variables can be written as
\begin{eqnarray}
s &=& m^2_{\chi_1}+m^2_e+2m_{\chi_1}(m_e+K_e)\,, \nonumber \\
t &=& m^2_{\chi_1}+m^2_{\chi_2}-2m_{\chi_1}(m_{\chi_2}+K_{\chi_2})\,.
\end{eqnarray}
The range,
$K^{\rm min}_{\chi_2} \leq K_{\chi_2} \leq K^{\rm max}_{\chi_2}$, is determined from~\cite{Jho:2020sku}
\begin{eqnarray}
\label{eq:range_t}
t=(p_1-p_3)^2=2m^2_e-
2\left(\sqrt{m^2_e+p^2_{\rm in}}\sqrt{m^2_e+p^2_{\rm out}}-p_{\rm in}p_{\rm out}\cos \theta_*\right)\,,
\end{eqnarray}
where the scattering angle in the center-of-mass frame takes values $0\leq \theta_* \leq \pi$. 
Here, $p_{\rm in}$ ($p_{\rm out}$) 
is  the momentum of the initial (final) state in the center-of-mass frame.
The cross section is given by
\begin{eqnarray}
\sigma_{\chi_1 e \to \chi_2 e}=
\int^{K^{\rm max}_{\chi_2}}_{K^{\rm min}_{\chi_2}}
dK_{\chi_2}~ \frac{2m_{\chi_2}}{16\pi \lambda(s,m^2_e,m^2_{\chi_1})}
\frac{1}{4} \sum |M|^2\,.
\end{eqnarray}
Note that in the static limit, justified by the non-relativistic velocity of the relic dark matter, 
the matrix element squared can be simplified to
\begin{equation}
 \frac14\sum |M|^2 = 
(\epsilon e g_\chi)^2
\frac{(2 m_e+\Delta m_\chi)^2(2 m_{\chi_1}+\Delta m_\chi)^2}
{(2m_e\Delta m_\chi + m^2_{A'})^2}\,.
\end{equation}
The $\chi_2$ flux at the Earth, assuming that it is produced isotropically in the Sun, is obtained from
\begin{eqnarray}
\label{eq:chi2_flux}
\frac{d \Phi_{\chi_2}}{dK_{\chi_2}} = 
 \frac{1}{4 \pi D^2_{\rm SE}} 
\frac{dC_c}{dK_{\chi_2}}\,,
\end{eqnarray}
where the distance between the Sun
and Earth, $D_{\rm SE} \equiv 1~{\rm AU}$. %
Subsequently, the $\chi_2$ interacts with an electron 
in the XENON1T detector. The event rate for the down scattering 
$\chi_2 e \to \chi_1 e$ is
\begin{eqnarray}
\frac{dR}{dK_r} = N_T^e \int \frac{d \Phi_{\chi_2}}{dK_{\chi_2}} 
\frac{d\sigma_{\chi_2 e \to \chi_1 e}}{dK_r}\,,
\end{eqnarray}
where $K_r$ is the electron recoil energy and  $N_T^e$ is the  
total effective number 
of  target electrons in the XENON1T detector.
Because only the outer shell electrons, from the 4$s$ to 5$p$ orbitals 
of Xe have a binding energy less than keV, 
we take 26 electrons for each Xe atom~\cite{Cao:2020bwd}.
We fold in the detector efficiency~\cite{Aprile:2020tmw} and energy resolution~\cite{Alonso-Alvarez:2020cdv} functions
to compute the event spectrum.

\begin{figure}[t]
\centering
\includegraphics[height=2.2in,angle=0]{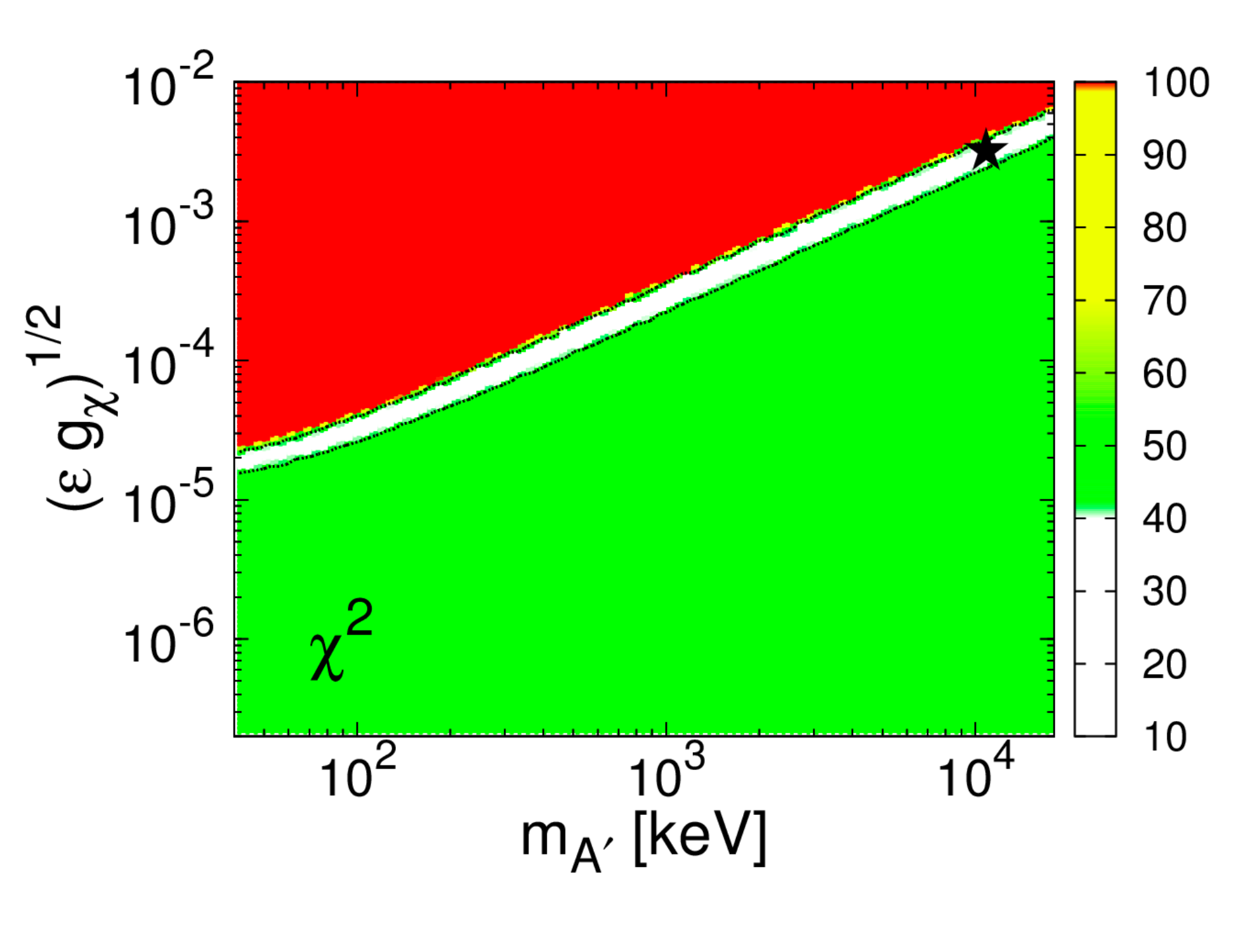}
\includegraphics[height=2.2in,angle=0]{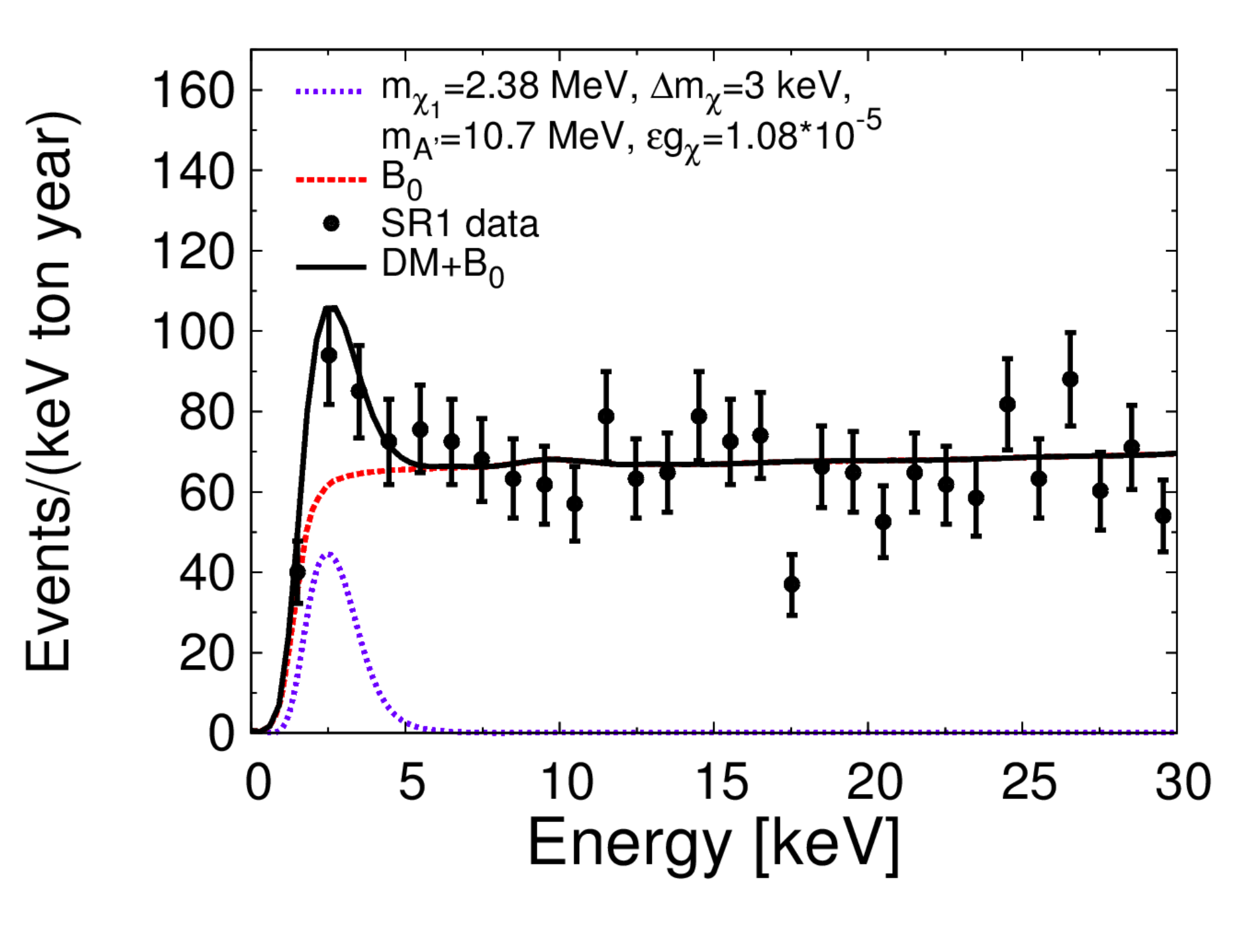}
\caption{\small \label{fig:BP-01}
{\boldmath{$m_{A'}> \Delta m_\chi$}}.
 The left panel shows the $\chi^2$ distribution from a fit to XENON1T data.
The $2\sigma$ contour corresponds to 
$\Delta \chi^2\equiv \chi^2- \chi^2_{\bf BP1}=6.17$, where $\chi^2_{\bf BP1}=35.76$
for the best-fit point {\bf BP1} in Eq.~(\ref{eq:sceanrio2_bestfit})
which is marked by a star. Its
electron recoil spectra is shown in the right panel.
}
\end{figure}

We fit the 29 bins of the XENON1T spectrum between 1~keV and 30~keV obtained during Science Run 1 (SR1) by scanning over $m_{\chi_1}, m_{A'}$ and $\epsilon g_\chi$, and find the best fit point,
\begin{eqnarray}
\label{eq:sceanrio2_bestfit}
\text{\bf BP1}: && (m_{\chi_1},\Delta m_\chi,m_{A'},\epsilon g_\chi)=(2.38\,{\rm MeV},3\,{\rm keV},10.7\,{\rm MeV},1.08\times 10^{-5})\,, 
\end{eqnarray}
with $\chi^2_{\bf BP1}=35.76$. The background-only hypothesis $B_0$~\cite{Aprile:2020tmw} has
$\chi^2_{\rm B_0}=46.69$.
The $\chi^2$ distribution in the $(m_{A'},\sqrt{\epsilon g_\chi})$  plane
is shown in the left panel of Fig.~\ref{fig:BP-01}, which displays
a strong correlation between $m_{A'}$ and $\epsilon g_\chi$. The spectrum of {\bf BP1} is shown in the right panel.

In principle, a light $A'$
can be thermally produced in the hot electron plasma of the Sun
via the coupling $\epsilon e$. Such a solar $A'$ flux can produce the recoil electrons responsible
for the XENON1T excess. However, for $m_{A'}> 40$~keV,
the $A'$ contribution is severely Boltzmann suppressed.

In Fig.~\ref{fig:scenario_2-1}, we overlay the XENON1T preferred regions 
for $g_\chi=\sqrt{4\pi}$ and $0.01$ 
with constraints from
HB stars~\cite{An:2013yfc,Redondo:2013lna},  fixed-target experiments~\cite{Harnik:2012ni} and
supernova 1987A~\cite{Chang:2018rso}. 
The unshaded region labelled ``SN1987A"  is an approximate representation of the region constrained by SN 1987A. 
 A detailed analysis is necessary to define the region precisely.
For $g_\chi=\sqrt{4\pi}$
the XENON1T preferred region is excluded, and for $g_\chi=0.01$,
the allowed window is
0.3~${\rm MeV}\lesssim m_{A'}\lesssim 1\,{\rm MeV}$.
%
%
However, the number of effective number of relativistic neutrinos at recombination, $N_{\rm eff}$,
is sensitive to $m_{A'}\lesssim 1$~MeV and $\epsilon \gg 10^{-8}$~\cite{Ibe:2019gpv}. 
The processes, 
$A'+\gamma \leftrightarrow e^- +e^+$ and 
$A'+e^\pm \leftrightarrow \gamma+ e^\pm$,
heat up only the electron-photon plasma and reduce $N_{\rm eff}$ to $\simeq 1.7$, which is 
excluded by cosmic microwave background data~\cite{Ibe:2019gpv}.

\begin{figure}[t]
\centering
\includegraphics[height=2.2in,angle=0]{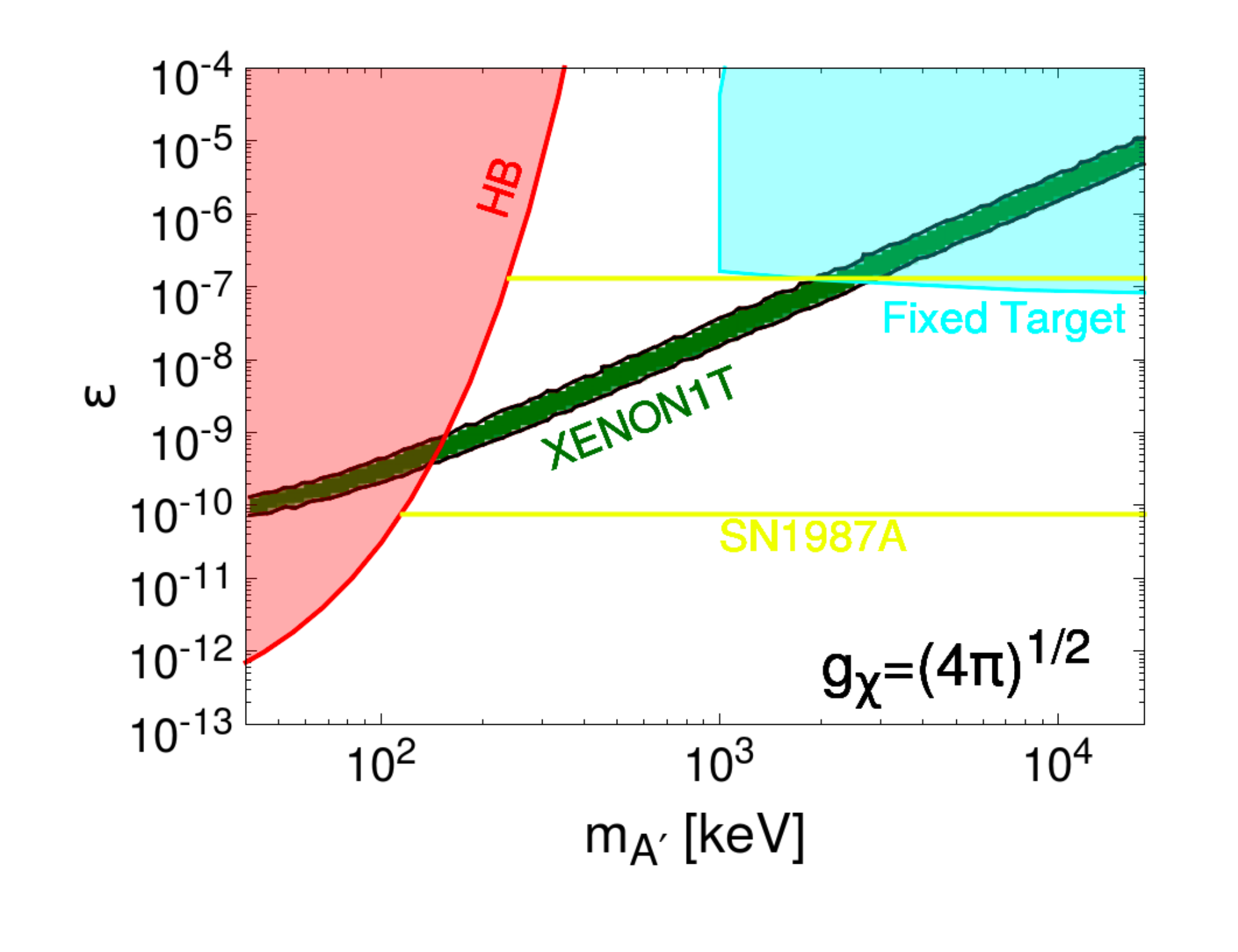}
\includegraphics[height=2.2in,angle=0]{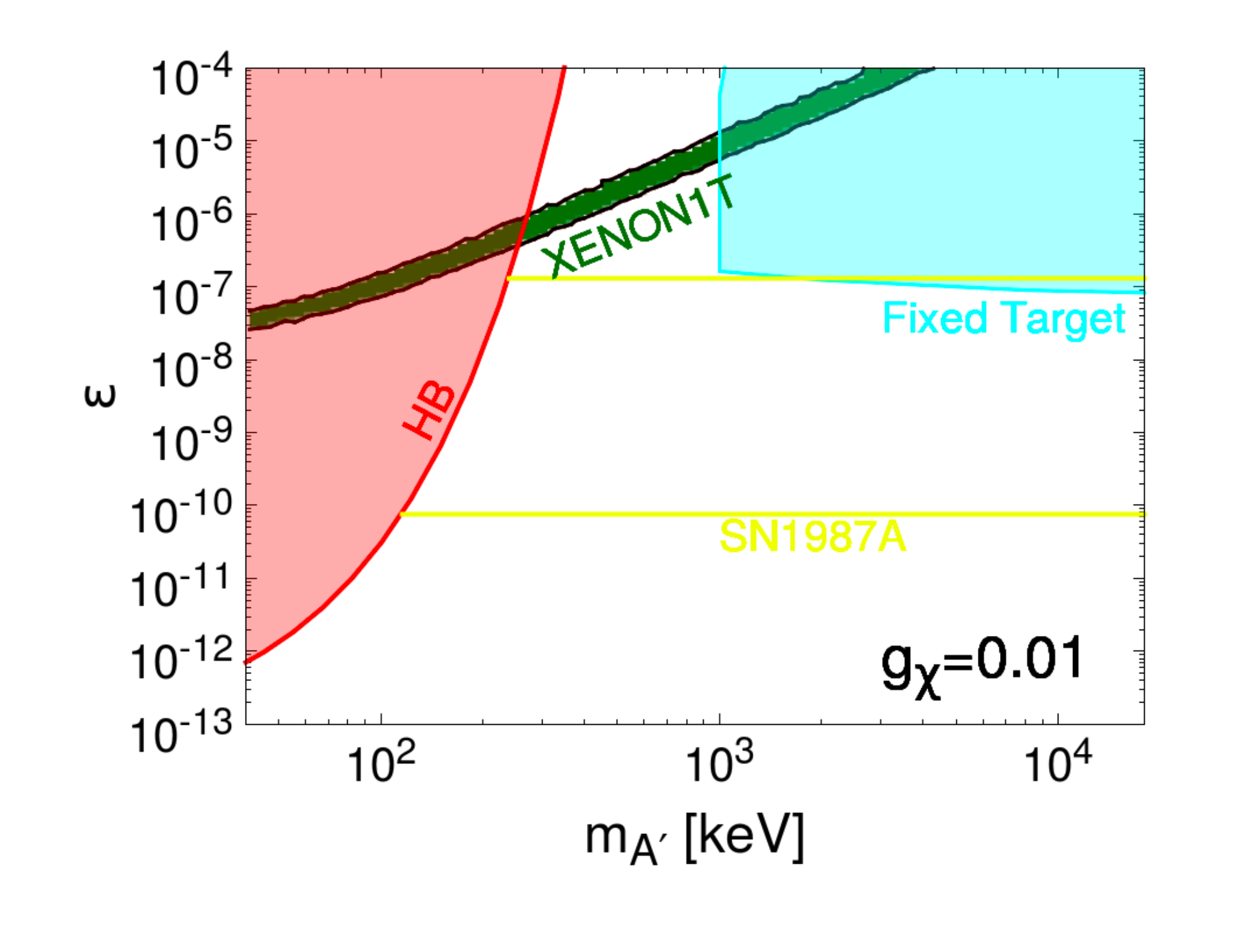}
\caption{\small \label{fig:scenario_2-1}
{\boldmath{$m_{A'}> \Delta m_\chi$}}.
The $2\sigma$ XENON1T allowed regions for $g_\chi=\sqrt{4\pi}$ and $g_\chi=0.01$ with $m_{\chi_1}=5$~MeV.
The constraints from HB stars~\cite{An:2013yfc,Redondo:2013lna}, 
 fixed-target experiments~\cite{Harnik:2012ni} and an approximate constraint from SN 1987A~\cite{Chang:2018rso} are also shown.
}
\end{figure}

The shape of the WDLF can be expressed as a simple power law if neutrino cooling is negligible~\cite{Giannotti:2015kwo}:
\begin{equation}
L_\gamma = 8.5\times10^{-4} L_\odot \left( \frac{T_{\rm WD}}{10^7\,{\rm K}} \right)^{3.5}
\ ,\label{eq:Lg}\end{equation}
where the solar luminosity is 
$ L_\odot = 2.39\times 10^{36}~{\rm GeV/s}$.
The exotic cooling rate of the WD can be similarly parametrized as
\begin{equation}
L_X=C_X L_\odot \left( \frac{T_{\rm WD}}{10^7\,{\rm K}} \right)^n\,.
\label{eq:Lx}\end{equation}
A fit to the data of Ref.~\cite{Bertolami:2014noa} yields
the best-fit values, $C_X=1.31\times 10^{-4}$ and $n=3.49$~\cite{Giannotti:2015kwo}. Taking the 
WD core temperature to be $10^7$~K $= 0.862$~keV, 
\begin{equation}
L_X\simeq  3.13\times 10^{32}~{\rm GeV/s}\,.
\label{rr}
\end{equation}
The relic DM $\chi$ can cool the WD through the process
$\chi_1 e\to \chi_2 e$. However, even at the  geometric limit
its rate is $2.2\times 10^{26}~{\rm GeV/s}$ for $m_{\chi_1}\simeq \mathcal{O}({\rm MeV})$ 
and $\Delta m_\chi\simeq \mathcal{O}({\rm keV})$, which is six orders of magnitude below
the required anomalous rate of Eq.~(\ref{rr}).

\subsection{\boldmath{$1~{\rm keV} < m_{A'}< \Delta m_\chi\,, 10~{\rm keV}$}}

We now consider $m_{A'} < \Delta m_\chi$ so that 
once $\chi_2$ is produced inside the Sun via 
$\chi_1 e \to \chi_2 e$, it decays promptly:  $\chi_2 \to A'\chi_1$.
Then, none of the $\chi_2$ propagate to Earth to generate a XENON1T signal.
The amplitude squared and decay width are given by
\begin{eqnarray}
\label{eq:amp_x2_ax1}
\frac{1}{2}\sum |M_{\chi_2 \to A'\chi_1}|^2 &=&
g^2_\chi 
\left\lbrace
m^2_{\chi_2}-6m_{\chi_1}m_{\chi_2}+m^2_{\chi_1}-2m^2_{A'}
+\frac{1}{m^2_{A'}} \left( m^2_{\chi_2}-m^2_{\chi_1}  \right)^2
\right\rbrace \nonumber \\
\Gamma_{\chi_2 \to A' \chi_1 }&=& \frac{1}{16\pi m_{\chi_2}} 
\frac{1}{2}\sum |M_{\chi_2 \to A' \chi_1}|^2
\lambda^{1/2}\left(1,\frac{m^2_{\chi_1}}{m^2_{\chi_2}},\frac{m^2_{A'}}{m^2_{\chi_2}} \right)\,.
\end{eqnarray}
The $(m^2_{\chi_2}-m^2_{\chi_1})^2/m^2_{A'}$ term 
in the amplitude squared
arises from the longitudinal component of $A'$, and dominates when $m_{A'}\ll \Delta m_\chi$.

The $A'$ easily escapes the Sun, propagates to Earth, and
contributes to the electron recoil signal at  XENON1T through
absorption by the bound electrons of Xe, $A' e
\to e$, analogous to the photoelectric effect.
The $A'$ absorption cross section per Xe atom is
\begin{eqnarray}
\label{eq:photoelectric}
\sigma_{A'}  \left\lbrace
\begin{array}{ll}
\simeq\left(\frac{m_{A'}}{\omega} \right)^2 
\epsilon^2 \sigma_{\gamma}\,\left( \frac{c}{v_{A'}} \right) &~~~~ \text{for longitudinal $A'$}\,,    \\
=\epsilon^2 \sigma_{\gamma}\,\left( \frac{c}{v_{A'}} \right) &~~~~ \text{for transverse $A'$~\cite{Pospelov:2008jk}}\,,
\end{array}
\right.
\end{eqnarray}
where $v_{A'}$ is the $A'$ velocity, and 
$\sigma_{\gamma}$ 
is the photoelectric cross section per Xe atom which is a function of the $A'$ energy $\omega$~\cite{Arisaka:2012pb,Fabbrichesi:2020wbt}. Note the
$(m_{A'}/\omega)^2$ suppression  for longitudinal modes.
Since $\chi_2$ with kinetic energy of $\mathcal{O}({\rm keV})$ 
is non-relativistic, $A'$ from $\chi_2$ decay has energy  
 $\omega\simeq \Delta m_\chi$. 
The $A'$ flux is identical to the $\chi_2$ flux and can be obtained from Eq.~(\ref{eq:chi2_flux}):
$$
\Phi^{\chi_2}_{A'}|_{\omega=\Delta m_\chi} = \int dK_{\chi_2}  \frac{d\Phi_{\chi_2}}{dK_{\chi_2}} \,.
$$
%

The $A'$ contribution from $\chi_2$ decay to the XENON1T event rate can be estimated as
follows.  Assume the up scattering  $\chi_1 e \to \chi_2 e$ in the Sun reaches
the geometric limit.  Then $\chi_2$ decay produces the maximum flux,
$\Phi^{\chi_2}_{A'}\simeq 7.44\times 10^4~{\rm cm^{-2} s^{-1}}$ for
$m_{\chi_1}\simeq \mathcal{O}({\rm MeV})$.  This yields an
electron recoil event rate $\simeq \epsilon^2 (2 \times 10^{21})~{\rm
  ton^{-1} yr^{-1}}$, which requires $\epsilon\simeq
\mathcal{O}(10^{-10})$ to explain the XENON1T excess.  %
The $(m_{A'},\epsilon)\simeq(\mathcal{O}(1 {\rm\,  keV}),
\mathcal{O}(10^{-10}))$ region of parameter space is ruled out by cooling 
constraints from the Sun and HB stars.

However, keV mass $A'$ are also produced
by the thermal plasma of the Sun. In fact, this contribution dominates the solar $A'$ flux.
We define the $A'$ number production rate 
per unit volume per unit energy as 
$\frac{d^2 \Gamma^{\rm \odot}_{A'}}{dVd\omega}$ which depends on the plasma frequency
$\omega_p=(n_e e^2/m_e)^{1\over2}$ with $n_e$  the number density of electrons. 
We will only be interested in $A'$ masses above the plasma frequency $\omega_p\simeq 0.3$~keV in the core of the Sun with $R_{\rm core}=0.18R_\odot$; in the center of HB stars, $\omega_p\simeq 2$~keV.
In this case, the number emission rate of longitudinal modes 
per unit volume per unit energy $\omega$ 
is given by~\cite{An:2013yfc,Redondo:2013lna}
\begin{eqnarray}
\label{eq:A_prod}
\frac{d^2\Gamma^{\rm \odot}_{A'}}{dVd\omega}\bigg|_L= 
\begin{array}{ll}
 %
\displaystyle\sum_{i=H, He} \frac{ 8Z^2_i \alpha^3 n_e n_{Z_i}}{3m^2_e}\,
\frac{\epsilon^2 m^2_{A'}}{\omega^4}\sqrt{\omega^2-m^2_{A'}}
\sqrt{\frac{8m_e}{\pi T}} f\left( \sqrt{\omega/T} \right)\,,~~
&  
\end{array}
\end{eqnarray}  
where $n_{Z_i}$ is the number density of ions of charge $-Z_i e$~\cite{Bahcall:1987jc},
and
\begin{eqnarray}
f(a)\equiv \int^{\infty}_a dx\, xe^{-x^2} 
\log \left| \frac{x+\sqrt{x^2-a^2}}{x-\sqrt{x^2-a^2}} \right|\,.
\end{eqnarray}
The corresponding expression for the transverse modes is~\cite{Redondo:2008aa}
\begin{eqnarray}
\frac{d^2\Gamma^{\rm \odot}_{A'}}{dVd\omega}\bigg|_T&=&
\frac{1}{\pi^2}\frac{\omega \sqrt{\omega^2-m^2_{A'}}}{e^{\frac{\omega}{T_\odot}}-1}
\frac{\epsilon^2 m^4_{A'}}{(\omega^2_p-m^2_{A'})^2+(\omega \Gamma_T)^2}
\Gamma_T \,,
 \nonumber \\
\Gamma_T &=& \frac{16 \pi^2 \alpha^3}{3m^2_e \omega^3}
\sqrt{\frac{2\pi m_e}{3T_\odot}} 
n_e \sum_{i=H,He}
Z^2_i n_{Z_i} \bar{g}_{i}
(1-e^{-\frac{\omega}{T_\odot}})
+ \frac{8 \pi \alpha^2}{3 m^2_e}n_e\,,
\end{eqnarray}
where $\bar{g}_{i}$ 
is a Boltzmann averaged Gaunt factor~\cite{Brussaard:1962zz}.
For $T_\odot = 1.15~{\rm keV}$ and $R_{\rm core}/R_\odot=0.18$, we find that setting $\bar{g}_{H}=\bar{g}_{He}=1$ reproduces the result of Ref.~\cite{Redondo:2013lna} in the $m_{A'}$ range of interest.

The $A'$ flux at Earth produced in the thermal plasma of the Sun is given by
\begin{eqnarray}
&& \frac{d\Phi^\odot_{A'}}{d\omega} = \frac{1}{ 4\pi\rm  D_{\rm SE}^2}\int dV 
\frac{d^2 \Gamma^{\rm \odot}_{A'}}{dVd\omega}\,.
\end{eqnarray}
The $A'$ contribution to the XENON1T spectrum is
\begin{eqnarray}
\label{eq:A_XENON1T_event}
\frac{dR}{d\omega} \simeq N_T\sigma_{A'} \frac{d\Phi^\odot_{A'}}{d\omega}\,,
\end{eqnarray}
where $N_T\simeq 4.52\times 10^{27}$ 
is the number of Xe atoms per ton.

It is interesting that the best fit value of $n$ in Eq.~(\ref{eq:Lx})
is the same as in Eq.~(\ref{eq:Lg}) for photon emission.
This implies that the emission of a light particle like $A'$
could provide the additional contribution to WD cooling.
In the core of a WD, $\omega_p \simeq 30$~keV. For $m_{A'}\ll \omega_p$, 
the resonant emission of longitudinal $A'$ is enhanced by $(\omega_p/m_{A'})^2$, 
can occur 
at any temperature, and is given by~\cite{An:2013yfc,Redondo:2013lna}:
\begin{eqnarray}
\label{eq:A_prodr}
\frac{d^2\Gamma^{\rm WD}_{A'}}{dVd\omega}= 
\frac{1}{4\pi}\frac{\epsilon^2 m^2_{A'}\omega^2}{e^{\omega/T}-1}\delta(\omega -\omega_p)\,.
&
\end{eqnarray}  
%
To explain the excess WD cooling by $A'$ emission,  
we require
\begin{eqnarray}
L_X &=& \int dV \int d\omega~ \omega \frac{d^2\Gamma^{\rm WD}_{A'}}{dVd\omega}
= \left(\frac{4}{3}\pi R^3_{\rm WD} \right) 
\frac{1}{4\pi} \frac{\epsilon^2 m^2_{A'}\omega^3_p}{e^{\omega_p/T_{\rm WD}}-1}\,, \nonumber \\
& {\rm i.e.,} & \ \ \ \ \epsilon\, m_{A'} \simeq 10^{-14}\, {\rm keV.} 
\label{band}
\end{eqnarray}
For $m_{A'} \ll \omega_p$, the region favored by the WDLF is a band with slope $-1$ in the $\log \epsilon-\log m_{A'}$ plane; see Fig.~\ref{fig:mA_mixing}. This is confirmed by numerical simulations up to $m_{A'} \simeq 1.6$~keV~\cite{Giannotti:2015kwo}. We have extended this band to 4~keV, above which the condition, $m_{A'}\ll \omega_p$, breaks down. For $2.5\,{\rm keV}\lesssim m_{A'}\lesssim 4\,{\rm keV}$, 
the WDLF favored region is consistent with all constraints. Note that Eqs.~(\ref{eq:A_XENON1T_event}) and~(\ref{band}) are independent of $g_\chi$.

\begin{figure}[t]
\centering
\includegraphics[height=2.2in,angle=0]{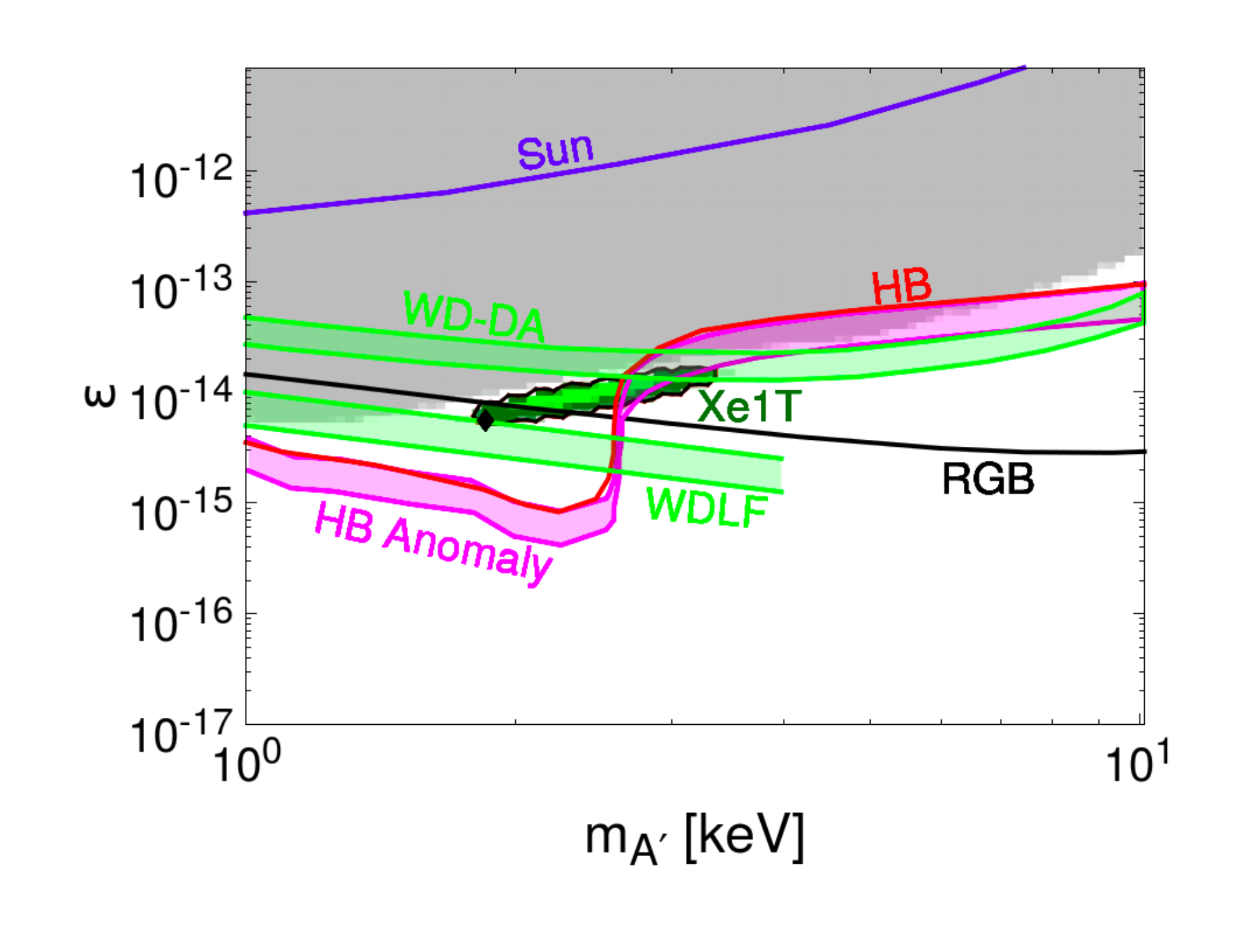}
\includegraphics[height=2.2in,angle=0]{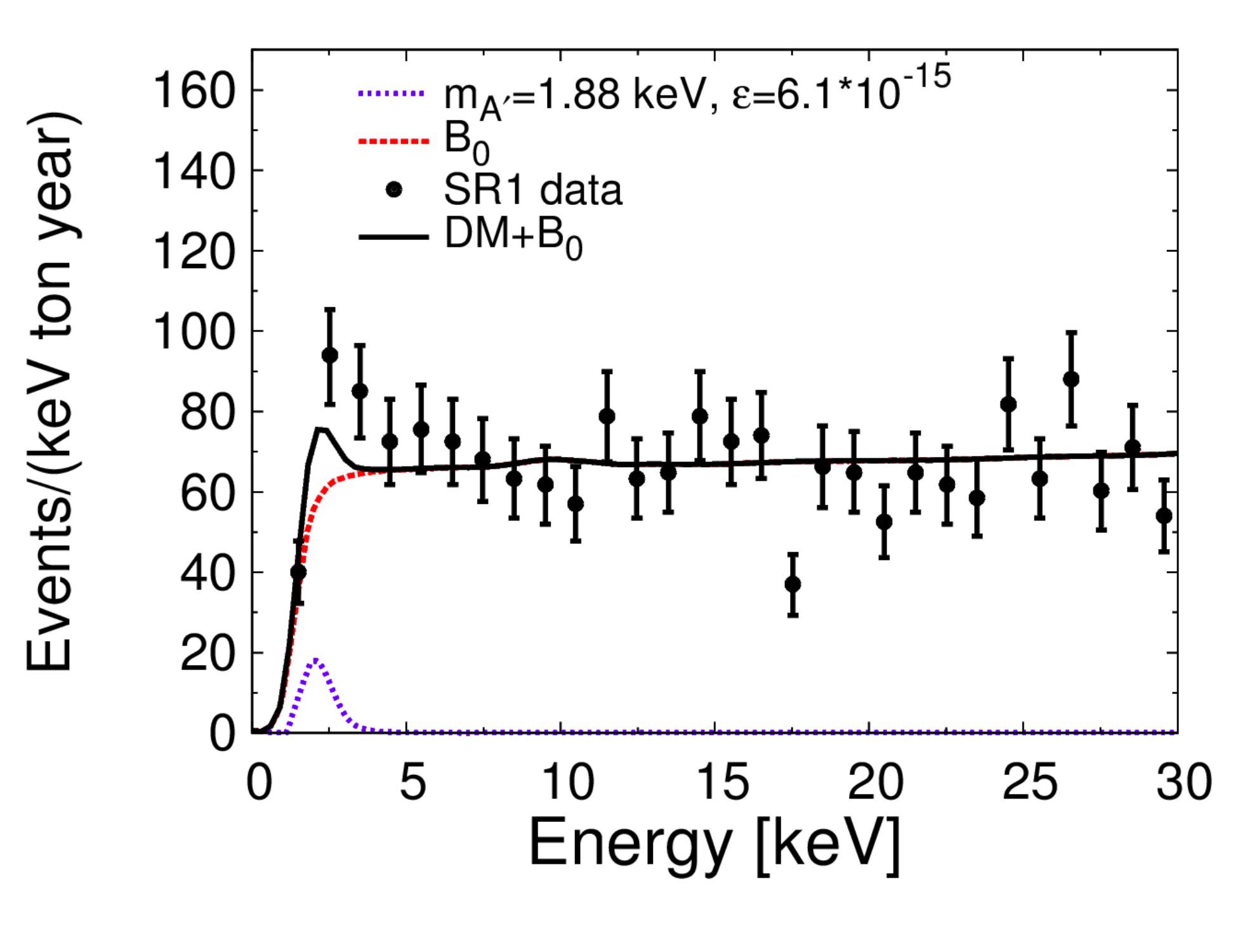}
\caption{\small \label{fig:mA_mixing}
{\boldmath{$m_{A'}< \Delta m_\chi$}}.
In the left-panel, we show the parameter regions favored by the cooling excesses suggested by the WDLF (at $2\sigma$), WDs of spectral type DA (WD-DA), and by the HB anomaly. The $1\sigma$ and $2\sigma$ allowed regions that explain the XENON1T are labelled ``Xe1T". 
Regions above the curves labelled, ``Sun'', ``HB'' (red) and ``RGB'' are excluded
at $2\sigma$ by the respective stellar constraints.
The gray shaded region 
labelled ``$\Delta\chi^2_{\rm Xe1T}\geq 100$''  is strongly disfavored.
The ``Xe1T" and ``WDLF" regions overlap in the vicinity of 
$(m_{A'},\epsilon)=(1.88~{\rm keV},6.1\times 10^{-15})$, marked by a diamond, and its electron recoil spectrum at XENON1T is shown in the right panel.
}
\end{figure}

In Fig.~\ref{fig:mA_mixing}, we show the $1\sigma$ and $2\sigma$ regions favored by XENON1T data and the $2\sigma$ region favored by the WDLF.  The $2\sigma$
regions barely overlap in the vicinity of
$$
{\bf BP2:}\ \ \ (m_{A'},\epsilon)=(1.88\,{\rm keV},6.1\times 10^{-15})\,,
$$
which has $\chi^2=42.12$ in a fit to the XENON1T data.
Its electron recoil spectrum is shown in the right panel. The best-fit point $(m_{A'},\epsilon)=(2.69~{\rm keV},1.26\times 10^{-14})$ with  
$\chi^2_{\rm best-fit}=36.07$ does not explain the WDLF data and is excluded by the RGB constraint.
The gray region marked ``${\rm \Delta\chi^2_{Xe1T}} \geq 100$'' 
is strongly disfavored because $\chi^2\geq 100$ for XENON1T data.

\begin{figure}[t]
\centering
\includegraphics[height=2.2in,angle=0]{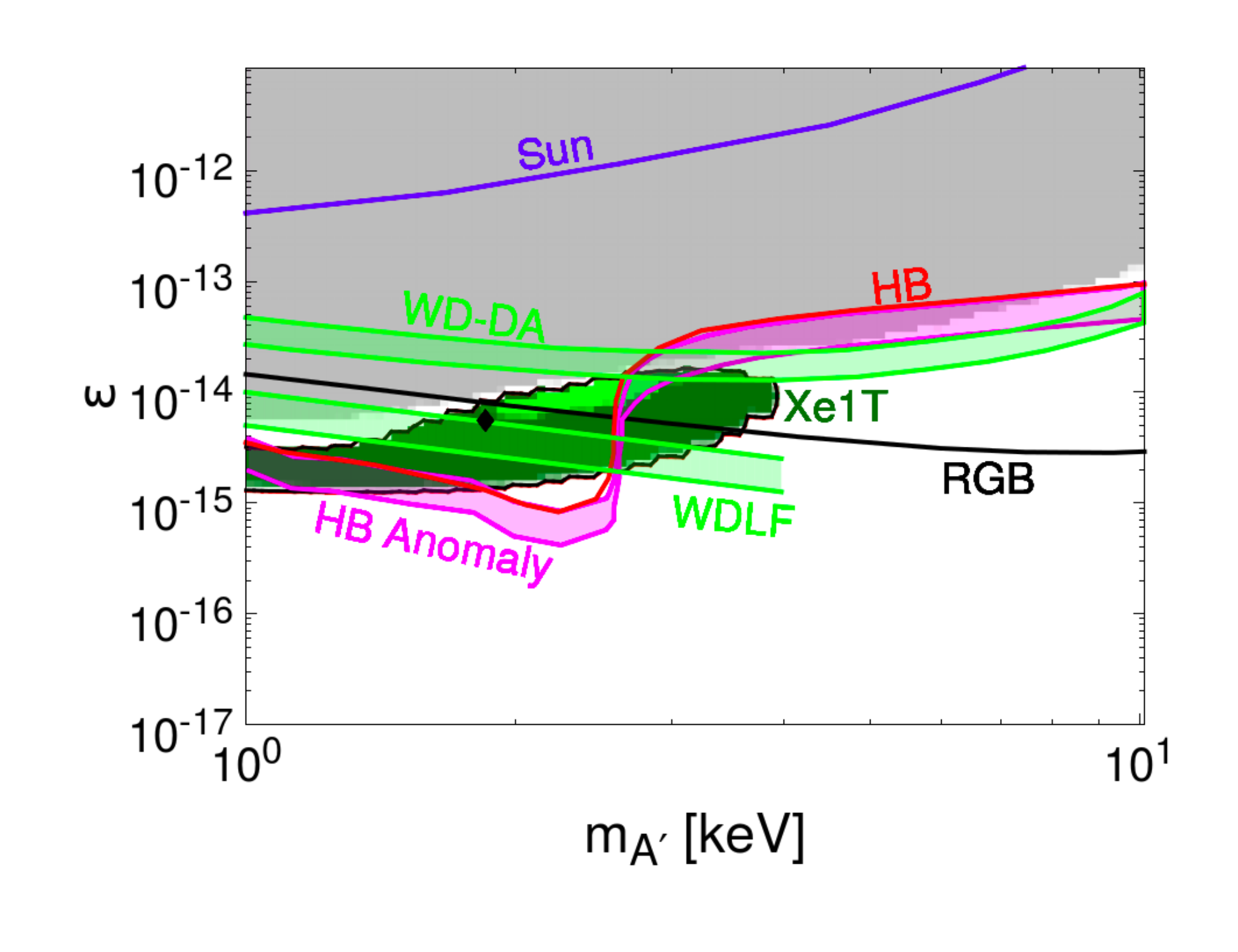}
\includegraphics[height=2.2in,angle=0]{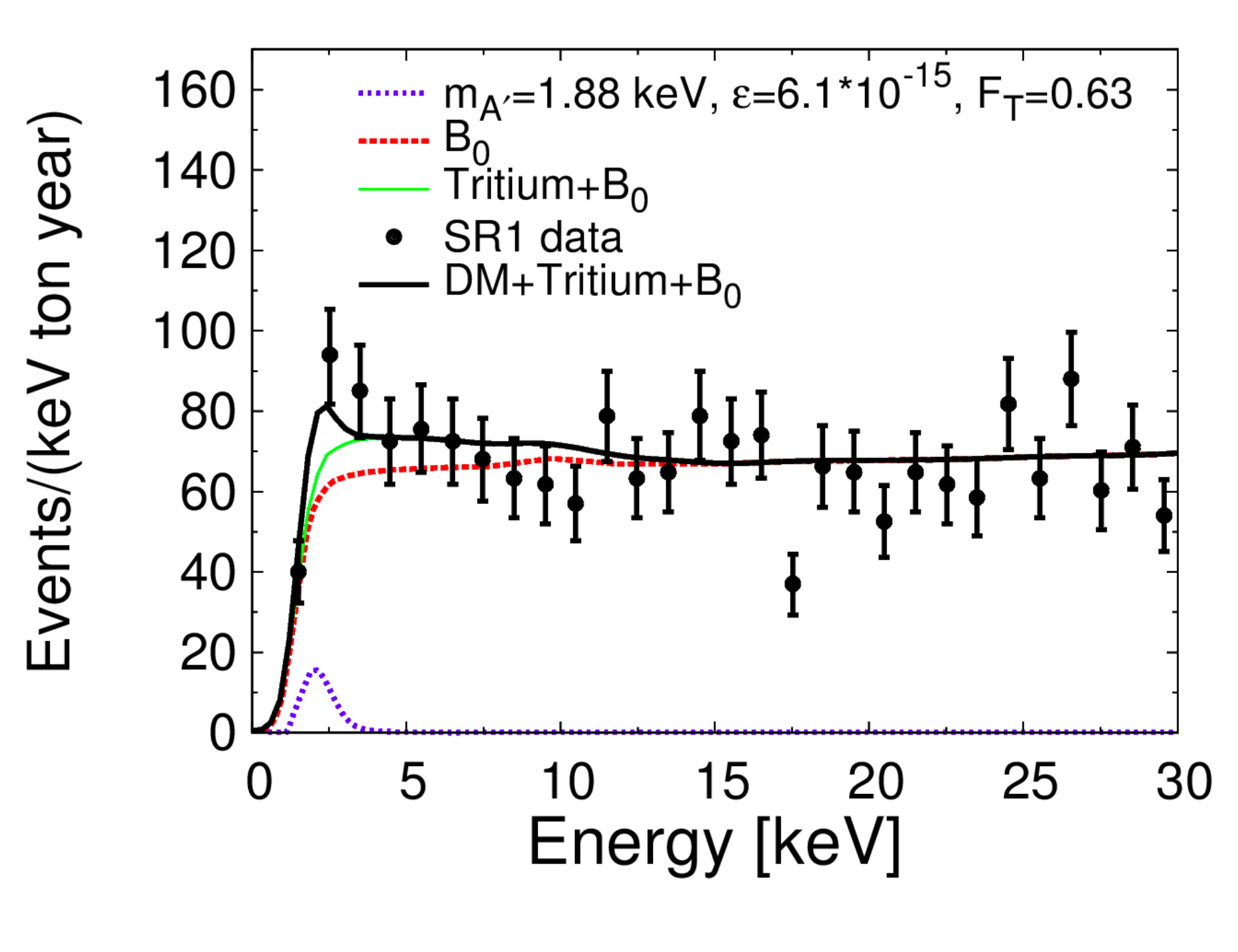}
\caption{\small \label{fig:mA_mixing_1}
Same as Fig.~\ref{fig:mA_mixing}, but including a tritium contribution
whose normalization factor $F_{\rm T}$ is 
a free parameter.
The right panel shows the spectrum of $(m_{A'},\epsilon,F_{\rm T})=(1.88~{\rm keV},6.1\times 10^{-15},0.63)$, a point
that explains the WDLF and XENON1T excesses.
}
\end{figure}

We therefore consider the possibility that a  tritium component contributes to the XENON1T excess in addition to solar $A'$. Then,
smaller values of $\epsilon$ are needed which could reconcile the WDLF excess. The result of allowing a free-floating normalization of the
$\beta$-decay contribution from tritium decay is shown in Fig.~\ref{fig:mA_mixing_1}. $F_{\rm{T}}=1$ sets the normalization of the tritium spectrum in Ref.~\cite{Aprile:2020tmw} and yields $\chi^2_{\rm T}=41.8$, which is close to $\chi^2_{\bf {BP2}}=42.12$. Including the tritium contribution with 
$F_{\rm T}=0.63$ improves the fit of {\bf BP2} to
$\chi^2_{\bf {BP2}}=39.36$. The expanded allowed region is even compatible with the HB anomaly.

In the scenario of Ref.~\cite{Alonso-Alvarez:2020cdv}, the hidden photon is the relic DM  candidate and its production
in the Sun is neglected.
Our preferred value of $\epsilon$ is larger than in Ref.~\cite{Alonso-Alvarez:2020cdv}
because the $A'$ fluxes differ.
For the DM $A'$,
the $A'$ flux is determined by the local DM number density 
$n_{\rm DM}\simeq 0.3~{\rm GeV/cm^3}$ 
and the dispersion velocity $v_{\rm DM}\simeq 10^{-3}c$, 
which gives 
$\Phi^{\rm DM}_{A'}(c/v_{\rm DM})\simeq 3\times 10^{15}~{\rm cm^{-2}s^{-1}}$.
The event rate 
$R^{\rm DM}_{A'}\simeq \epsilon^2_{\rm DM} 
(\Phi^{\rm DM}_{A'}/v_{\rm DM})
N_T\sigma_{\gamma}c$ yields $(m_{A'},\epsilon_{\rm DM})=(2.8\,{\rm keV},8.6\times 10^{-16})$ 
as the best fit point~\cite{Alonso-Alvarez:2020cdv}.
To compare  with our model, 
we fix $m_{A'}=2.8$~keV, multiply the solar $A'$ differential flux 
at the Earth 
by $c/v_{A'}$ and integrate over energy. 
Both scenarios should give the same event rate at XENON1T:
\begin{eqnarray}
&& \epsilon^2_{\rm DM} \left( \frac{\Phi^{\rm DM}_{A'}c}{v_{\rm DM}} \right)
= \epsilon^2 \int d\omega \frac{d\Phi^\odot_{A'}}{d\omega} 
\left( \frac{c}{v_{A'}} \right)  \simeq  
\left(\frac{\epsilon}{10^{-7}} \right)^4 
\times 1.67\times 10^{13}~{\rm cm^{-2}s^{-1}}\,.
\end{eqnarray}
Note that the event rate $ \sim \epsilon^4$ in our model because the event rate picks up  two powers of $\epsilon$ each at $A'$  production in the Sun 
and its interaction in the detector. An order of magnitude larger value,  $\epsilon=1.15\times 10^{-14}$,
is preferred in our scenario. 

\subsection{\boldmath{$\Delta m_\chi < m_{A'}< 40~{\rm keV}$}}

In the previous two subsections either $\chi_2$ or $A'$ produced the XENON1T signal, but not both.
Here, we study a hybrid case in which both $\chi_2$ and $A'$ produce signals in XENON. This can
occur if $\Delta m_\chi < m_{A'}$ so that $\chi_2$ decay is forbidden, and $m_{A'}$ is light enough that it can be produced in the Sun.
Because we set $\Delta m_\chi =3$~keV, these conditions are satisfied for $3~{\rm keV}< m_{A'}< 40~{\rm keV}$.
  The parameter region from an analysis of XENON1T data up to 50~keV is shown in Fig.~\ref{fig:hybrid_case}. A tritium contribution is not included.
  A fit of the background to the 38 bins gives $\chi^2_{B_0}=56.06$.

\begin{figure}[t]
\centering
\includegraphics[height=3.1in,angle=0]{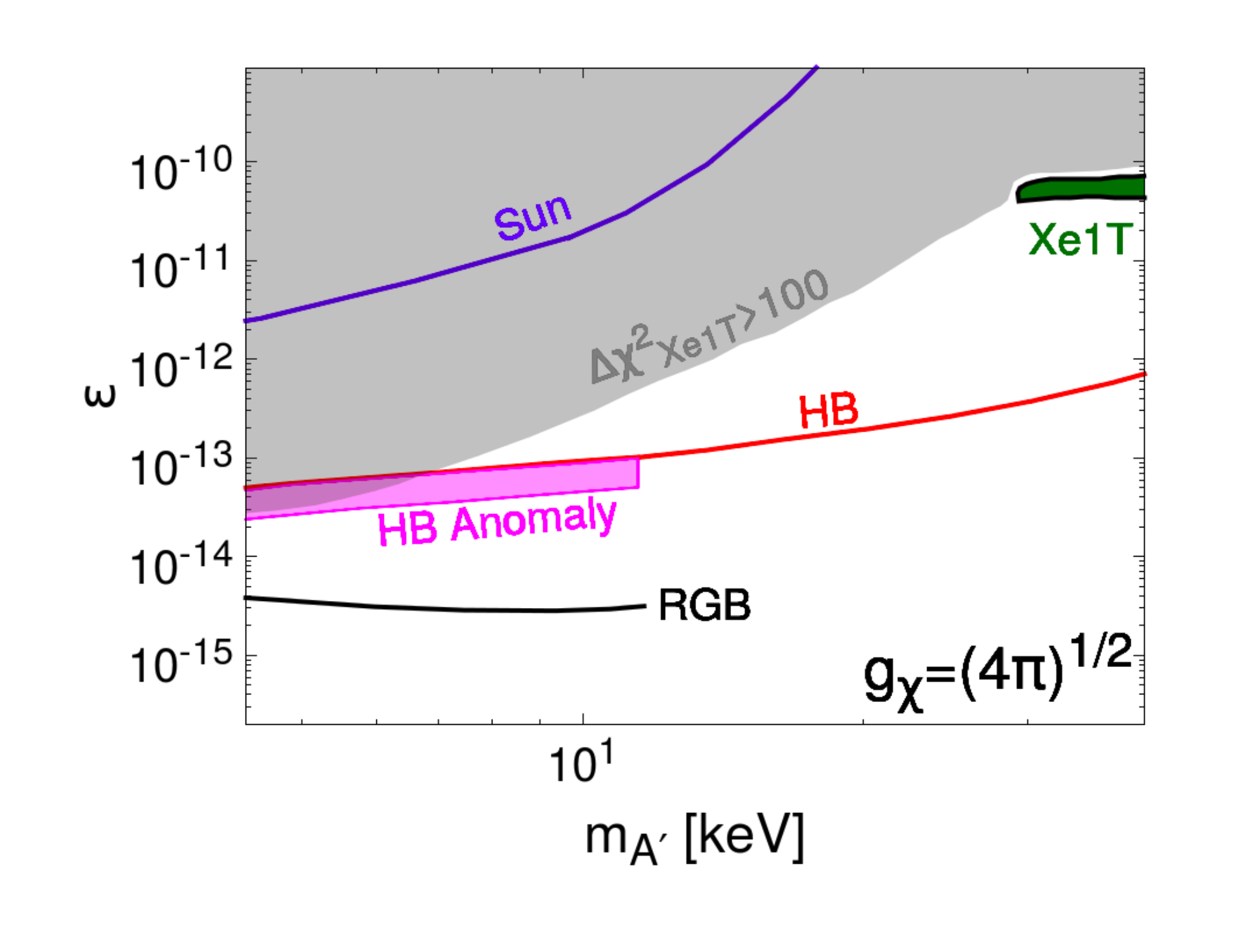}
\caption{\small \label{fig:hybrid_case}
{\boldmath{$\Delta m_\chi =3~{\rm keV} < m_{A'}< 40~{\rm keV}$}}. Similar to Fig.~\ref{fig:mA_mixing}, but in this hybrid case, 
both $\chi_2$ and $A'$ produce signals in XENON. In this analysis we include XENON1T data up to 50~keV and set $m_{\chi_1}=5\,{\rm MeV}$. 
}
\end{figure}

\subsubsection{Double-peak signal at XENONnT}
\label{sec:XENONnT}

\begin{figure}[h]
\centering
\includegraphics[height=2.2in,angle=0]{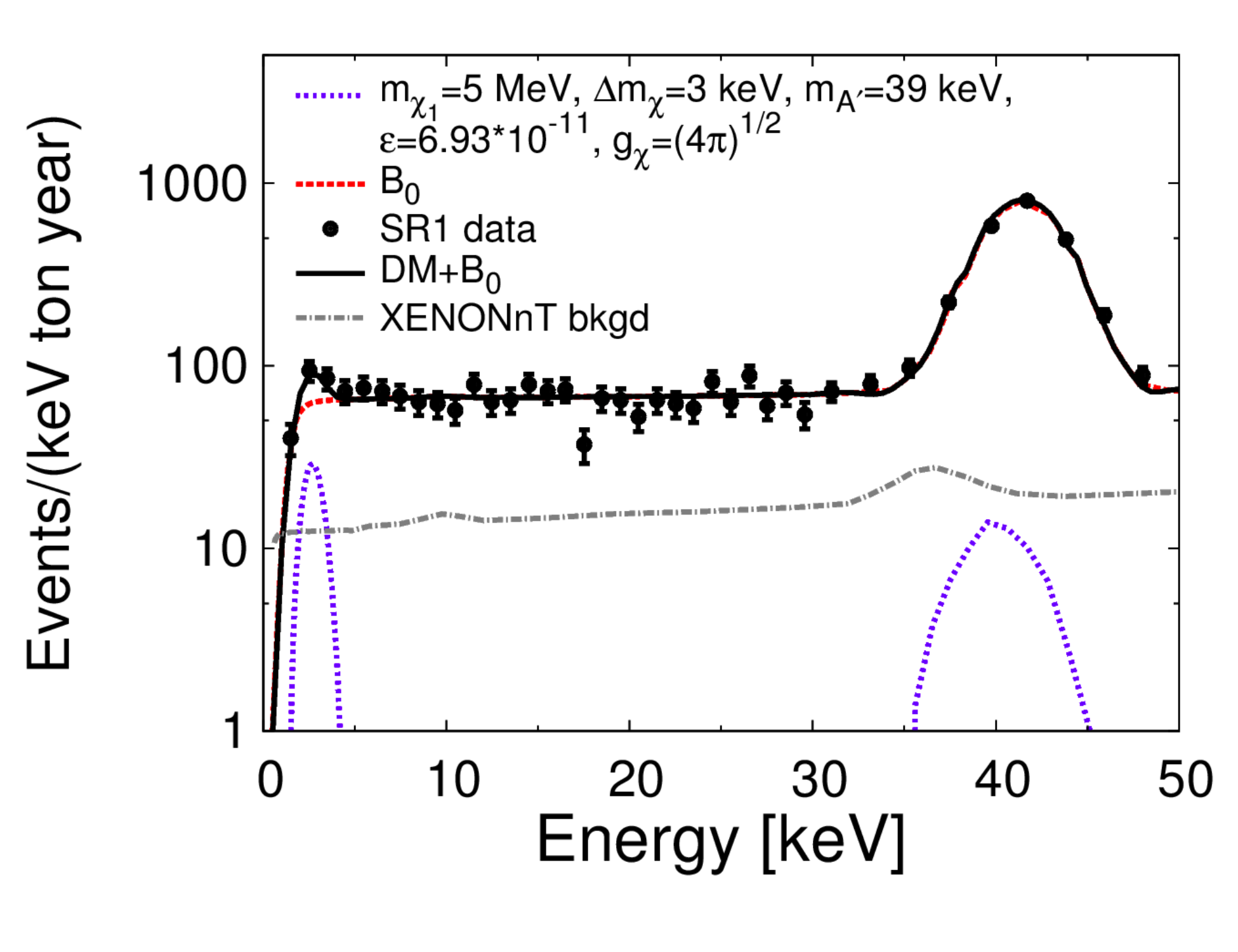}
\includegraphics[height=2.2in,angle=0]{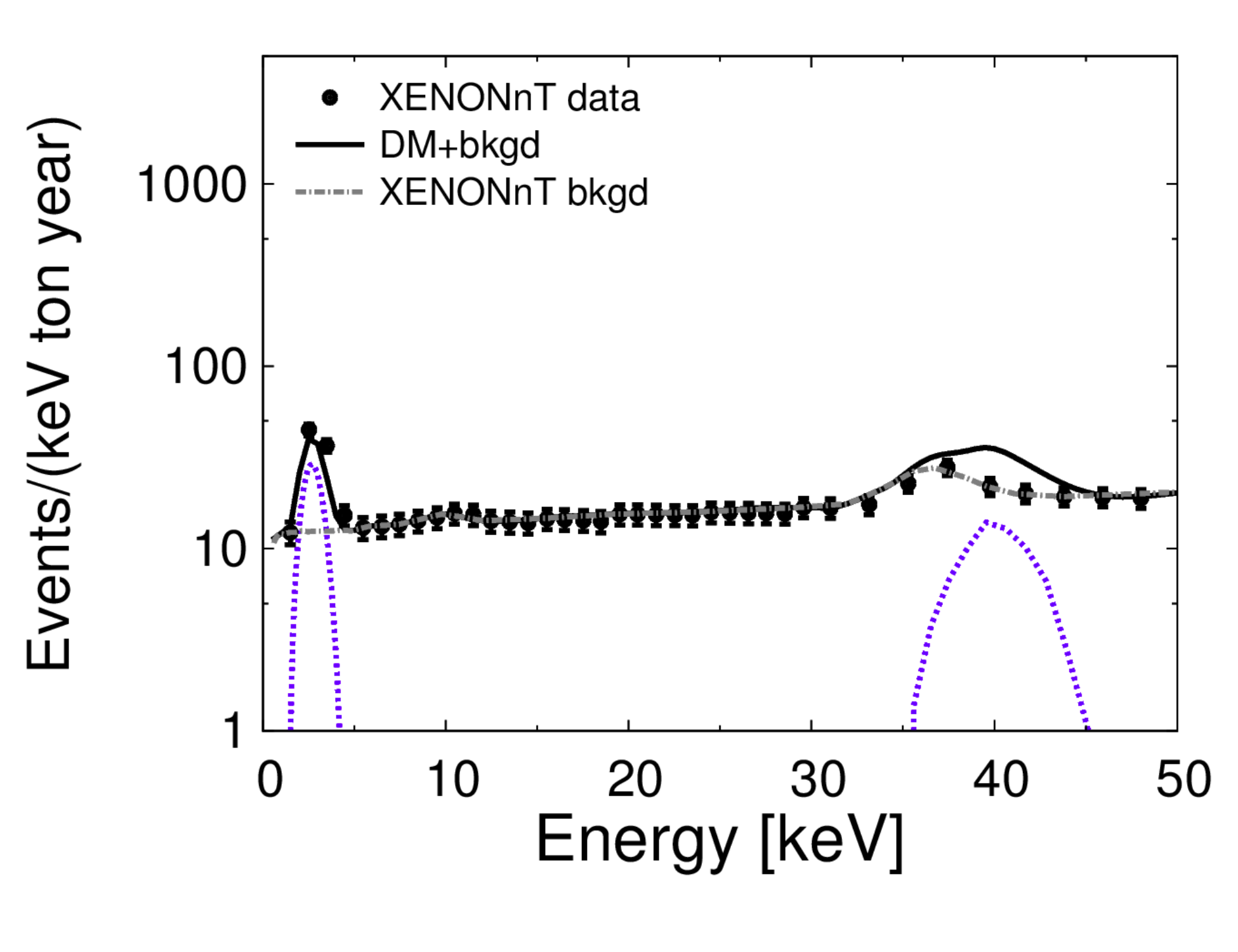}
\caption{\small \label{fig:combine_doublepeak} 
The double-peak spectra for the hybrid case.
The left panel shows the XENON1T spectrum, and the right panel shows that both peaks
will be evident in 4 ton-year of XENONnT data.
}
\end{figure}

The point
$(m_{A'},\epsilon,g_\chi)\simeq (39~{\rm keV},6.93\times 10^{-11},\sqrt{4\pi})$ is interesting 
because the two contributions from $\chi_2$ and $A'$  to XENON1T data
are  comparable. This point has $\chi^2=46.22$, i.e.,
$\Delta \chi^2=9.84$, which for 3 parameters gives a $p$-value of 0.020,
so it is allowed at 98\%~CL. A
distinctive double peak  in the electron recoil spectrum is predicted, as illustrated 
in the left panel of Fig.~\ref{fig:combine_doublepeak}.
The peak at 3~keV is from the narrow mass gap of $\chi$,
and the peak around 39~keV is due to the solar $A'$ of the same mass value.
The signal at 39~keV is 
hidden in the huge background induced 
by the neutron-activated isotope $^{131m}Xe$~\cite{Aprile:2020tmw}.
 Note that this parameter  region does not explain the WDLF anomaly and
is excluded by HB cooling constraints.

To evaluate the sensitivity of XENONnT data~\cite{Aprile:2020vtw}, 
we assume the same number of excess events per keV-ton-year 
as XENON1T with uncertainties reduced by a factor of 2.5
for a 4 ton-year exposure; see the right panel of Fig.~\ref{fig:combine_doublepeak}.
We also assume that xenon purification will remove tritium to an undetectable level.
For the parameter point in the left panel, the signal gives $\chi^2=48.66$ and the background hypothesis has $\chi^2_{\rm bkgd}=159.59$, so the signal will be easily detected.

Next, we suppose that XENONnT does not see an excess, and the background expectation~\cite{Aprile:2020vtw}  in the recoil energy range, 1~keV to 13~keV, 
is consistent with data.
Then XENONnT data
will be able to exclude the XENON1T allowed region of Fig.~\ref{fig:mA_mixing}, which did not assume a tritium contribution; 
see the left panel of Fig.~\ref{fig:XENONnT}.
From the right panel we see that XENONnT will not be able to exclude the scenario that allows for a tritium component to the XENON1T excess.

\begin{figure}[t]
\centering
\includegraphics[height=2.2in,angle=0]{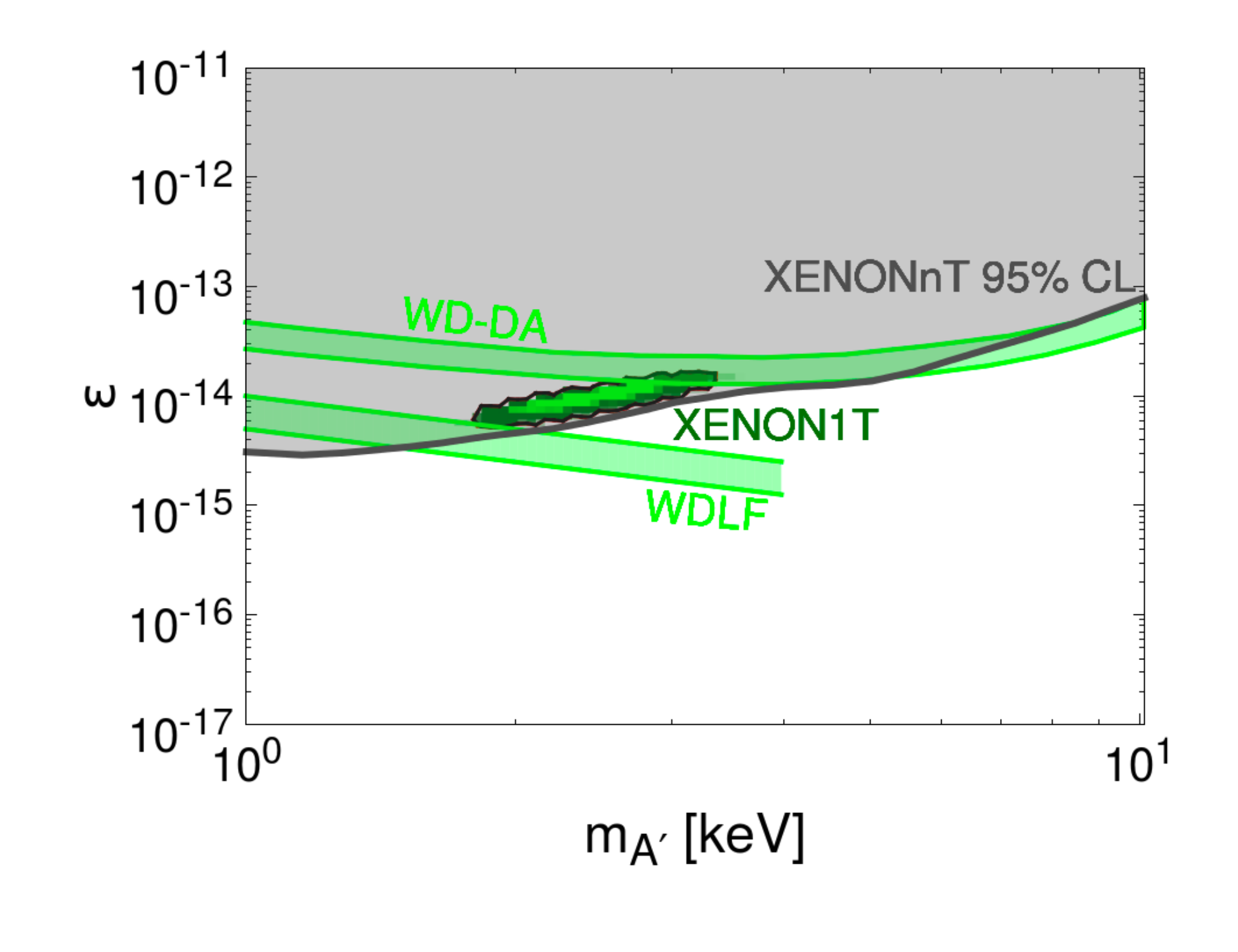}
\includegraphics[height=2.2in,angle=0]{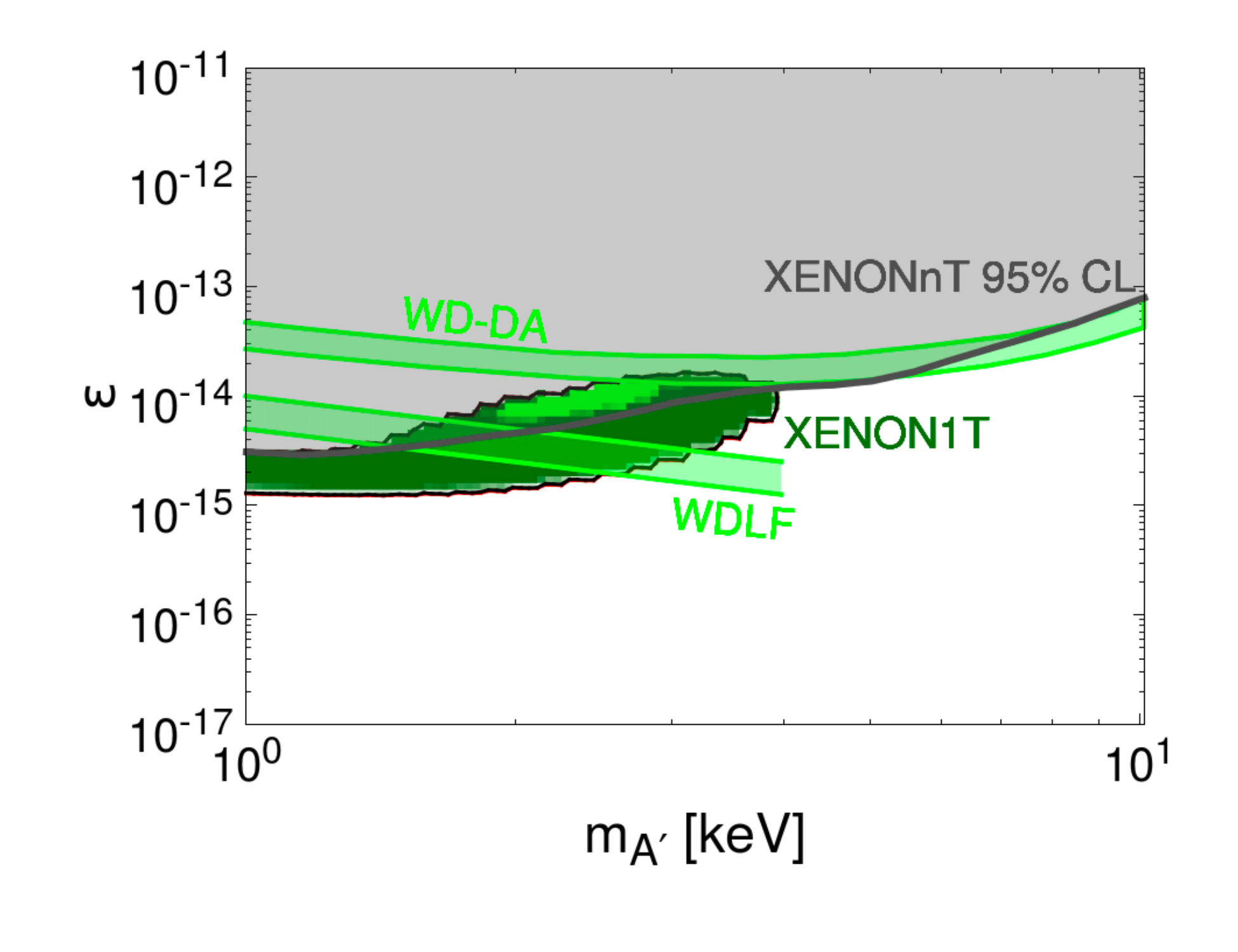}
\caption{\small \label{fig:XENONnT}
The expected 95\% exclusion limit from XENONnT data
between 1~keV and 13~keV assuming no excess is seen in 4 ton-year of XENONnT data. The XENON1T regions in the left (right) panel 
are from Fig.~\ref{fig:mA_mixing} (Fig.~\ref{fig:mA_mixing_1}) without (with) a tritium component to the excess.
}
\end{figure}

\bigskip

\section{Summary}
\label{sec:conclusion}

We proposed dark-photon mediated inelastic DM as an explanation
of the WDLF cooling anomaly as well as
the XENON1T electron recoil excess.
The entire DM halo is composed
of the lighter dark particle $\chi_1$,
and thermal electrons inside the Sun provide a power source
to excite $\chi_1$ into $\chi_2$ through the process, $\chi_1 e\to \chi_2 e$.
For $m_{A'} \gsim \Delta m_\chi\,, 40$~keV, the stable $\chi_2$
propagates to Earth and contributes to electron recoil events at XENON1T via the down scattering process, $\chi_2 e \to \chi_1 e$.
The recoil spectrum features a peak around the mass difference
between $\chi_1$ and $\chi_2$, $\Delta m_\chi = 3~{\rm keV}$.

For $m_{A'}< \Delta m_\chi\,, 10$~keV, the $\chi_2$ produced in the Sun
decays promptly to $A' \chi_1$
so that $A'$  contributes, though subdominantly,
to the XENON1T signal.
The dominant $A'$ production in the Sun arises from the thermal plasma which can explain the XENON1T excess with
$(m_{A'},\epsilon)=(1.88\,{\rm keV},6.1\times 10^{-15})$ provided a tritium component also contributes to the XENON1T spectrum.

For $\Delta m_\chi < m_{A'} < 40$~keV, a distinctive signature of our model is a
second peak in the XENON1T spectrum that is currently not detectable.
However, the parameter region where this occurs does not explain the white dwarf cooling excess and is severely constrained by HB cooling.

If the same event excess is found at XENONnT, it will be definitively confirmed.
On the other hand, if XENONnT data are consistent with the background, the parameter region favored by the XENON1T anomaly (without a tritium contribution) will be ruled out at $2\sigma$.
The preferred region common to the XENON1T and WDLF anomalies
can not be fully probed by  XENONnT after complete removal of tritium; see Fig.~\ref{fig:XENONnT}.

\def\dum{

\section{Conclusion}
\label{sec:conclusion}

We proposed the inelastic DM with the vector mediator $A'$, which  
couples to the electron,
in order to explain the WDLF anomaly, as well as 
the XENON1T electron recoil excess.
The entire DM halo is assumed to be composed 
of the lighter dark particle $\chi_1$,
and then the thermal electron inside the Sun is treated as a power source
to excite $\chi_1$ into $\chi_2$ through the process $e\chi_1 \to e \chi_2$.
For $m_{A'} > \Delta m_\chi$ case, the stable $\chi_2$ 
propagates to Earth and contributes to electron recoil at XENON1T
via the down scattering process $\chi_2 e \to \chi_1 e$.
Result in the recoil spectrum features a peak around the mass difference 
between $\chi_1$ and $\chi_2$, here we choose $\Delta m_\chi= 3~{\rm keV}$.

For $m_{A'}< \Delta m_\chi$, the produced $\chi_2$ from the Sun
would decay promptly into $A' + \chi_1$
such that the decay product $A'$  contributes, though subdominantely, 
to XENON1T signal.
The dominant $A'$ production is from the thermal plasma of the Sun,
which can explain the XENON1T excess with 
$(m_{A'},\epsilon)=(2.69{\rm keV},4.47\times 10^{-15})$.
That is different from the hidden photon scenario 
of Ref.\cite{Alonso-Alvarez:2020cdv}, 
where they identified hidden photon as dark matter, 
but did not include the $A'$ production from the Sun.
In fact, our XENON1T preferred region requires larger value of $\epsilon$
than that in Ref.\cite{Alonso-Alvarez:2020cdv},
which makes it closer to the WDLF region.
Within $2\sigma$ C.L., there is overlapping 
between the WDLF and XENON1T regions on the $(m_{A'},\epsilon)$ plane.

Distinctive signature from our proposed model 
can generate double peak spectrum 
at XENON1T detector at 
$(m_{A'},g_e,g_\chi)\simeq (34{\rm keV},2\times 10^{-11},\sqrt{4\pi})$.
However, this parameter region is severely constrained by HB cooling.

The XENON1T anomaly preferred region 
$(m_{A'},\epsilon)=(2.69{\rm keV},4.47\times 10^{-15})$
can be fully probed by the future XENONnT.
Specifically, if the same event excess be found at XENONnT, it will be confirmed upto $3\sigma$.
On the other hand, if the result is consistent with background expectation at XENONnT, 
it will rule out the $1\sigma$ region of XENON1T anoamly.
}

\section*{Acknowledgements}  
D.M. is supported in
part by the U.S. DOE under Grant No. de-sc0010504.
P.T is supported by National Research Foundation of Korea (NRF) funded by the Ministry of Education, Science and Technology (NRF-2020R1I1A1A01066413).
\appendix


\end{document}